\documentclass[sigconf]{acmart}

\usepackage{graphicx}
\usepackage{pgfplots}
\usepackage{booktabs}
\usepackage{makecell}
\usepackage{multirow}
\usepackage{subcaption}
\usepackage[inline]{enumitem}
\usepackage{siunitx}
\usepackage{tikz}
\usetikzlibrary{calc,trees,positioning,arrows,chains,shapes.geometric,%
  decorations.pathreplacing,decorations.pathmorphing,shapes,%
  matrix,shapes.symbols,plotmarks,decorations.markings,shadows}

\setcopyright{none}

\newcommand{\ignore}[1]{}

\def\Hk{\makebox[1.6mm][l]{\hspace{0.2mm}\footnotesize k}}

\newcommand\Pstation[1]{\textsf{``#1''}}

\makeatletter
\DeclareRobustCommand*\cal{\@fontswitch\relax\mathcal}
\makeatother

\DeclareMathOperator*{\psimi}{\mathit{sim}_{\scriptscriptstyle \text{P}}}

\DeclareMathOperator*{\edsimi}{\mathit{sim}_{\scriptscriptstyle \text{ED}}}

\newcommand{\ngram}[1]{\texttt{#1}}

\definecolor{darkblue}{rgb}{0.0, 0.0, 0.55}
\definecolor{darkgreen}{rgb}{0.0, 0.55, 0}

\newcommand{\tikzcircle}[2][red,fill=red]{\tikz[baseline=-0.75ex]\draw[#1,radius=#2] (0,0) circle ;}%

\renewcommand\footnotetextcopyrightpermission[1]{}
\begin{document}
\title{Similarity Classification of Public Transit Stations}

\author{Hannah Bast}
\affiliation{%
  \institution{University of Freiburg}
  \city{Freiburg}
  \state{Germany}
}
\email{bast@cs.uni-freiburg.de}

\author{Patrick Brosi}
\affiliation{%
  \institution{University of Freiburg}
  \city{Freiburg}
  \state{Germany}
}
\email{brosi@cs.uni-freiburg.de}

\author{Markus N\"ather}
\affiliation{%
  \institution{University of Freiburg}
  \city{Freiburg}
  \state{Germany}
}
\email{naetherm@cs.uni-freiburg.de}

\begin{abstract}
  We study the following problem: given two public transit station identifiers $A$ and $B$, each with a label and a geographic coordinate, decide whether $A$ and $B$ describe the same station.
  For example, for "St Pancras International" at $(51.5306, -0.1253)$ and "London St Pancras" at $(51.5319, -0.1269)$, the answer would be "Yes".
  This problem frequently arises in areas where public transit data is used, for example in geographic information systems, schedule merging, route planning, or map matching.
  We consider several baseline methods based on geographic distance and simple string similarity measures.
  We also experiment with more elaborate string similarity measures and manually created normalization rules.
	Our experiments show that these baseline methods produce good, but not fully satisfactory results.
  We therefore develop an approach based on a random forest classifier which is trained on matching trigrams between two stations, their distance, and their position on an interwoven grid.
  All approaches are evaluated on extensive ground truth datasets we generated from OpenStreetMap (OSM) data: (1) The union of Great Britain and Ireland and (2) the union of Germany, Switzerland, and Austria.
  On all datasets, our learning-based approach achieves an F1 score of over $99\%$, while even the most elaborate baseline approach (based on TFIDF scores and the geographic distance) achieves an F1 score of at most $94\%$, and a naive approach of using a geographical distance threshold achieves an F1 score of only $75\%$.   
  Both our training and testing datasets are publicly available\footnote{\url{https://staty.cs.uni-freiburg.de/datasets}}.
\end{abstract}

\maketitle

\section{Introduction}

\begin{figure}
    \centering
    \includegraphics[trim=0 11 0 0, clip, width=0.48\textwidth]{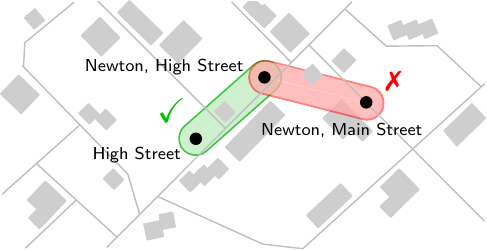}
    \caption{Station similarity classification problems in the fictional town of Newton. Each colored area marks a classification problem between two stops. Bus stop \Pstation{High Street} should be classified as similar to \Pstation{Newton, High Street} (green). The latter should be classified as not similar to \Pstation{Newton, Main Street} (red).}
    \label{FIG:intro}
\end{figure}

A recurring problem with public transit data is to decide whether two station identifiers, both consisting of a label and a geographic position, describe the same real-world station.
Figure~\ref{FIG:intro} gives an example of three (ficticious) station identifiers within a distance of 100 meters.
While it is obvious for humans that \Pstation{Newton, High Street} and \Pstation{High Street} describe the same station, but \Pstation{Newton, High Street} and \Pstation{Newton, Main Street} are different, it is nontrivial to decide automatically.
This has ramifications in various areas where station disambiguation is an important preprocessing step:

\textbf{GIS.} In the context of geographic information systems, search queries for \Pstation{London St Pancras} might exclude a station labeled \Pstation{St Pancras International} if the two stations are not disambiguated.

\begin{figure}[b]
    \centering
    \includegraphics[width=0.42\textwidth]{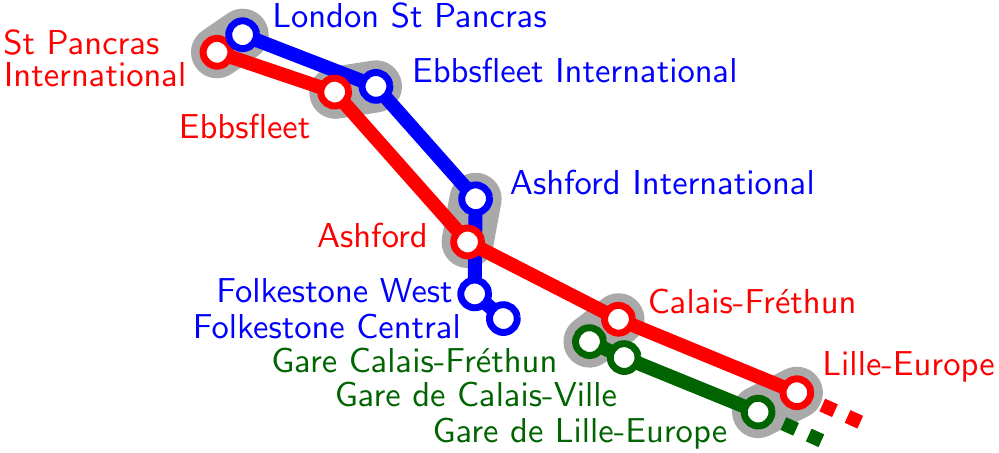}
    \caption{Three (simplified) schedule datasets for national \tikzcircle[black, fill=darkgreen]{3pt}, international \tikzcircle[black, fill=red]{3pt} and regional \tikzcircle[black, fill=blue]{3pt} trains. The station identifier pairs encircled in gray describe the same real-world station, but their labels and positions differ per dataset.}
    \label{FIG:intro2}
\end{figure}

\textbf{Schedule Merging.} When multiple schedule datasets are merged, for example to create a uniform regional dataset consisting of multiple agencies, station identifiers must be properly disambiguated.
In Figure~\ref{FIG:intro2}, a regional train schedule (blue) contains a station identifier \Pstation{London St Pancras}, but an international schedule dataset (red) identifies the same station by a label \Pstation{St Pancras International} at a slightly different location.

\textbf{Route Planning.} A route planner fed with schedule datasets without proper disambiguation might e.g. display an unnecessary footpath between \Pstation{London St Pancras} and \Pstation{St Pancras International} for routes changing trains at St Pancras.
This might both confuse travelers and compromise the cost metric.
If the route planner does not compute footpaths, it might also happen that the route cannot be found at all.

\textbf{Map-Matching.} When map-matching is done with stations as sample points, a station point labeled \Pstation{London St Pancras} and positioned at the station entrance might not be correctly matched to a station in the geo-spatial data labeled \Pstation{St Pancras International Station} and positioned on the tracks.

\begin{figure*}[t]
    \centering
    \includegraphics[trim=0 50 0 0, clip, width=0.475\textwidth]{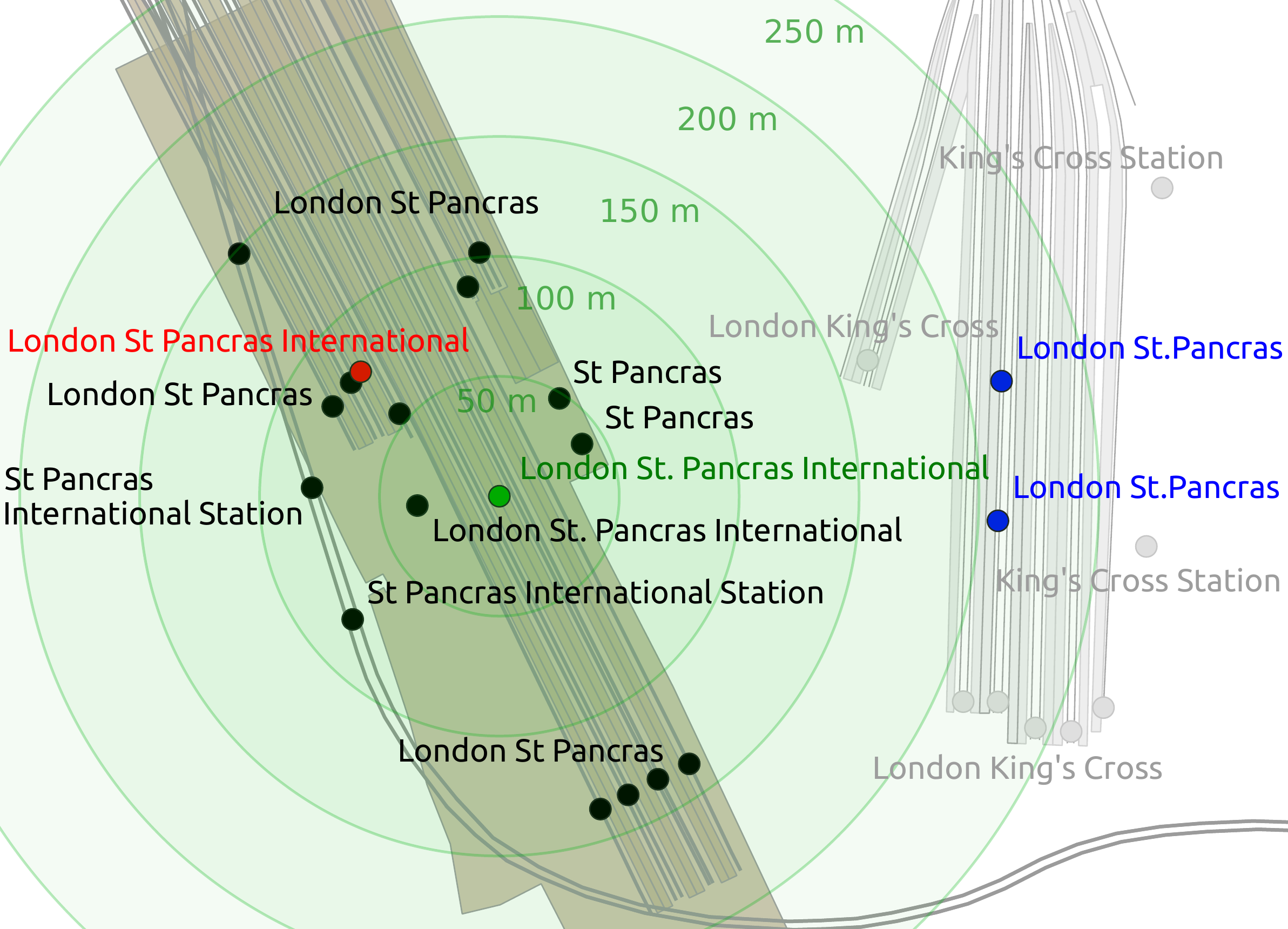}
    \hfill
    \includegraphics[trim=0 50 0 0, clip, width=0.475\textwidth]{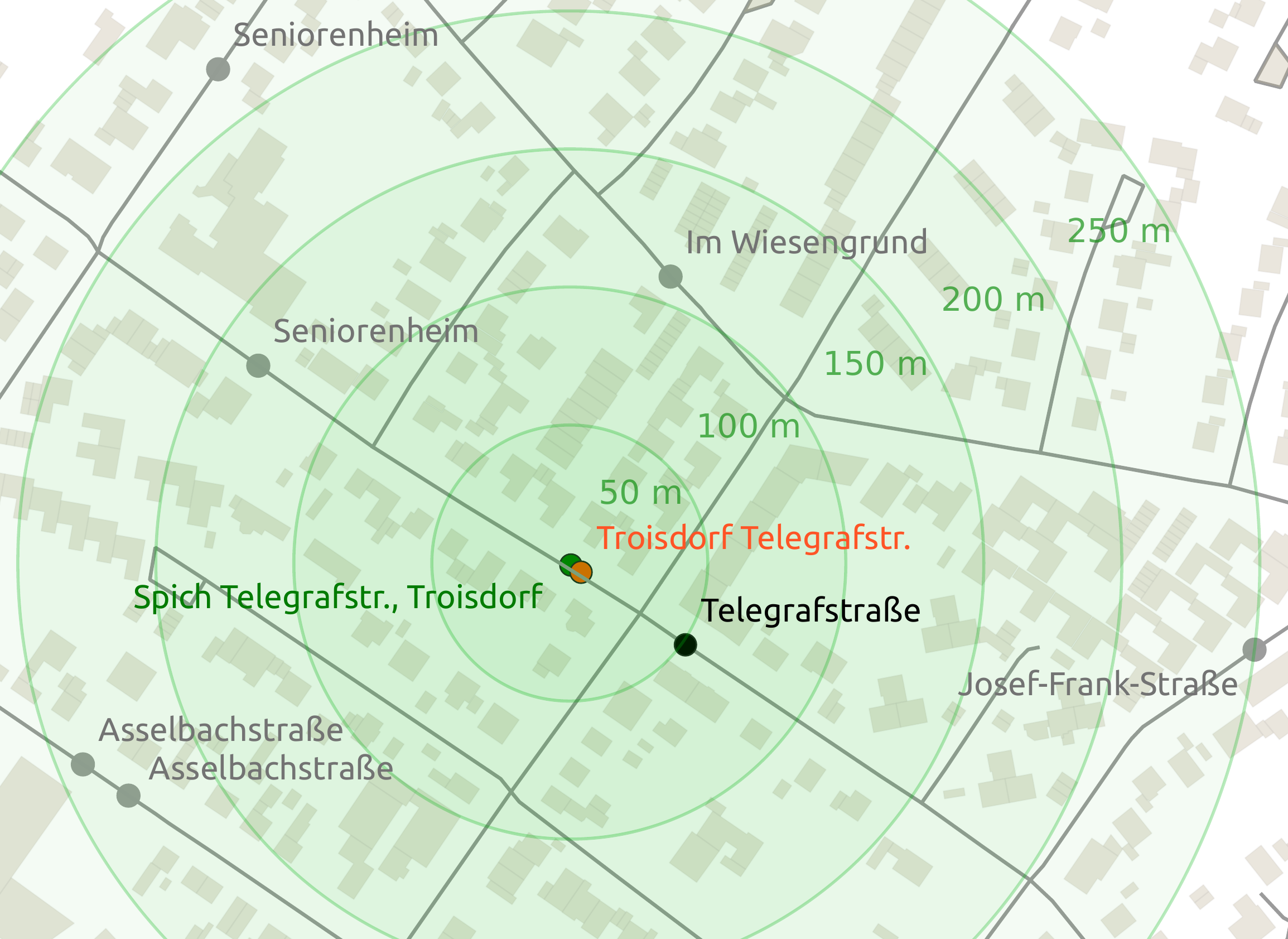}
    \vspace{-.5em}
    \caption{Left: Station identifiers for London St Pancras as they appear in three different datasets: OpenStreetMap (\tikzcircle[black, fill=black]{3pt}~OSM,~black), Deutsche Bahn schedule (\tikzcircle[black, fill=green]{3pt}~DB,~green), Association of Train Operating Companies schedule (\tikzcircle[black, fill=red]{3pt}~ATOC,~red), EuroStar schedule (\tikzcircle[black, fill=blue]{3pt}~ES,~blue). Note the distance of over 200 meters from the \tikzcircle[black, fill=green]{3pt}~DB station to the \tikzcircle[black, fill=blue]{3pt}~ES station. Also note that the nearest station in the OSM dataset for both \tikzcircle[black, fill=blue]{3pt}~ES stations is King's Cross station. Right: Identifiers for the bus stop \Pstation{Telegrafstra\ss e} in Troisdorf, Germany. \tikzcircle[black, fill=black]{3pt}~OSM identifiers are again given in black, identifiers from the local transit authority schedule \tikzcircle[black, fill=orange]{3pt}~VRS in orange.}
    \label{FIG:example}
    \vspace{-.5em}
\end{figure*}

Figure~\ref{FIG:example} gives two real-world examples of the challenges.
The goal of this work is to find robust approaches to this problem.
We start with a formal problem definition in Section~\ref{SEC:probdef} and discuss the characteristics of station positions and labels in Section~\ref{SEC:characteristics}.
In Section~\ref{SEC:classification}, we give an overview over several baseline similarity measures between station identifiers.
We then develop a learning-based approach which trains a random forest classifier on pairs of similar and non-similar stations.
Section~\ref{SEC:evalsetup} then describes how we obtained ground truth data from OpenStreetMap (OSM).
All approaches are evaluated in Section~\ref{SEC:results}.

\vspace{-4pt}
\subsection{Contributions}
\label{SEC:contrib}

We consider the following as our key contributions:
\begin{itemize}[topsep=0pt, parsep=0.5mm,leftmargin=0mm,itemindent=4mm,itemsep=-1pt]
\renewcommand\labelitemi{$\bullet$}
    \item We study the characteristics of station identifiers belonging to the same station based on multiple international datasets.
    \item We evaluate several baseline classification techniques based on geographic distance and/or various string similarity measures, including a novel measure called the Best Token Subsequence Similarity (BTS).
    \item We describe a learning-based approach that uses a random forest classifier, trained (among other features) on the difference of trigram occurrences in both identifiers and their positions on an interwoven geographic grid.
    \item We evaluate all techniques on datasets covering Germany, Switzerland, Austria, Great Britain and Ireland. On our largest dataset, our learning-based approach achieves an F1 score of over 99\%, while the baseline approaches achieve an F1 score of at most 94\%.
\end{itemize}

\subsection{Problem Definition}
\label{SEC:probdef}

A station identifier $s$ is a triple $(n, \phi, \lambda)$, where $n$ is the station name (for example, \Pstation{London St. Pancras}) and $\phi$ and $\lambda$ are the latitude and longitude of its position, respectively.
Our goal is to find a function $c$ that maps pairs of stations to $\{0, 1\}$ such that $c(s_a, s_b) = 1$ when $s_a$ and $s_b$ belong to the same real-world station, and $c(s_a, s_b) = 0$ otherwise.
We will refer to $c$ as a \emph{classifier}.

We will design and evaluate functions based on explicitly constructed similarity measures, as well as parametrized functions, where we learn the parameters from training data.

We obtain our ground truth from stations in \texttt{public\_transport=} \texttt{stop\_area} relations in OSM.
In a nutshell, according to the criteria for this relation\footnote{\url{https://wiki.openstreetmap.org/wiki/Tag:public\_transport\%3Dstop_area}},
two OSM nodes belong to the same such relation if they are both part of a station that is commonly presented as a single unit to passengers.
For example, if a large train station consists of multiple tracks, the OSM nodes describing the tracks are considered pairwise similar.
If a bus stop serves two directions, the platforms of the two directions are considered similar.

\subsection{Related Work}
\label{SEC:related}

Our work is closely related to previous work on string label similarity and similarity measures for geographic locations, as they are for example used for Point of Interest (POI) matching.

String label similarity classification is a recurring problem in various fields of research.
For example, in~\cite{bilenko03}, similarity measures between short database records (e.g. city names or first and/or last names) were investigated, among them the Jaro and Jaro-Winkler similarity and token-based measures like TFIDF scores or the Jaccard index.
Similarity measures for name-matching of generic entities were for example evaluated in~\cite{cohen03}.
Another area of research where label similarity is of interest is author disambiguation.
For example, in~\cite{han04}, author names of scientific publications were disambiguated by training a Naive Bayes classifier or a support-vector machine.
For recent surveys on author name disambiguation techniques, see for example \cite{ferreira12} and \cite{hussain17}.

In the area of Geographic Information Retrieval, similarity measures for geographic locations try to rank geographic locations (often combined with some labels) with respect to a textual user query (for example ``Bar in Vienna'').
In \cite{jones01}, such a measure based on a geographic ontology which represents spatial relationships between locations was described.
In \cite{andrade06}, a similar measure was combined with BM25 scores for measuring textual similarity.

The closely related field of POI matching tries to find POI pairs which describe the same real-world location, often to merge geo-spatial datasets \cite{safra10}.
This is typically done via a combination of spatial and textual similarity measures.
For example, in \cite{scheffler12} a binary spatial similarity measure based on a threshold for the Euclidean distance was combined with a two-phased approach which first considered the edit distance, and if no match was found, the TFIDF similarity.
In \cite{liu14}, the goal was to match spatio-textual data (consisting of a geographic location and a textual description) as it appears for example in social media to real-word POIs, and receive the top-$k$ best matches.
For the spatial similarity, a normalized Euclidean distance was used.
For the textual similarity, the weighted Jaccard index was chosen (as a weight, the inverse document frequency (IDF) was proposed).
A recent work \cite{deng19} assumes that the geographic distances between matching POIs follow an exponential distribution and models the spatial similarity measure accordingly.
Additionally, a label similarity (based on the edit distance), an address similarity (based on TFIDF scores) and a category similarity (based on hierarchical category data that was part of the input) were considered.
For a recent overview over existing work on POI matching, also see \cite{deng19}.

To the best of our knowledge, the applicability of such methods to station identifiers has not been investigated so far (it is also not obvious that they should work, because of the special nature of station identifiers, see Section~\ref{SEC:characteristics}).
We evaluate the similarity measures typically found in this area in Section~\ref{SEC:results}.

In \cite{bast18}, a station label similarity measure called token subsequence edit distance was described to improve map-matching results for schedule data, but without offering a thorough evaluation.
We evaluate an improved variant of this measure (called BTS).

Our method of encoding geographic positions on an interwoven grid is reminiscent of recent work on positional encoding for machine learning (see e.g. \cite{gehring17}).
For example, in \cite{vaswani17}, sequence token positions were mapped to sinusoidal functions of different frequencies to allow learning of both absolute and relative position characteristics.

As our ground truth dataset is generated from OSM data, our work is also related to previous work that applied machine learning approaches to OSM data.
For example, in \cite{funke15}, missing road data was extrapolated by training a classifier which decided whether a road segment should be present between two candidate nodes.
The authors of \cite{karagiannakis15} trained an SVM on OSM data to recommend categories for newly inserted OSM nodes.
In \cite{funke17}, a random forest classifier trained on $k$-grams of amenity names was used to infer missing tags (e.g. the \texttt{cuisine} tag for restaurants).
In \cite{bast20}, we trained a random forest classifier on OSM station data to automatically correct public transit station tagging, but without giving a thorough evaluation of the underlying classification results.

\section{Station Identifier Characteristics}
\label{SEC:characteristics}

As mentioned above, for two different station identifiers belonging to the same real-world station, geographic coordinates may differ significantly and labels may differ greatly.
In this section, we describe characteristics of both the labels and coordinates of station identifiers as well as the rationale behind different labeling and positioning variants.

\subsection{Characteristics of Geographic Positions}
\label{SEC:poscharacteristics}

There are mainly three reasons why similar station identifiers have inconsistent coordinates: (1) different principles guiding the placement, (2) imprecise coordinates, and (3) human error.
We again consider Figure~\ref{FIG:example},~left.
The OSM station identifiers for \Pstation{St Pancras} are placed either directly on the tracks, at station entrances, or somewhere around the station centroid (which is not well-defined).
All three approaches are reasonable.
Even worse, the station identifiers for \Pstation{St Pancras} from the EuroStar dataset are located in the middle of \Pstation{King's Cross} station (it is hard to tell whether this is a human error or a precision problem).
In Figure~\ref{FIG:example},~right, it is equally hard to tell whether the different locations for \Pstation{Telegrafstra\ss e} are caused by precision problems or human error.

\begin{figure}
    \begin{center}
        \resizebox {0.48\textwidth} {!} {
            \input{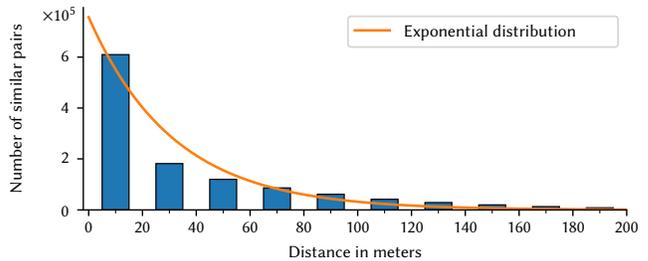}
        }
    \end{center}
    \vspace{-1em}
    \caption{Distribution of geographic distances between (unique) similar station identifier pairs in our OSM-based ground truth dataset for Germany, Austria and Switzerland.}
    \vspace{-3pt}
    \label{FIG:disthisto}
\end{figure}

We examined the distribution of distances between stations marked as similar in our OSM ground truth data for Germany, Austria and Switzerland.
The results can be seen in Figure~\ref{FIG:disthisto}.
While most of the similar station pairs (69\%) were within a distance of 50 meters, 19\% were between 50 and 100 meters apart, and 12\% were over 100 meters apart.
It appears that the distances between similar pairs roughly follow an exponential distribution.

\subsection{Characteristics of Station Labels}
\label{SEC:labelcharacteristics}

Figure~\ref{FIG:labelfreiburg} gives an example of different label variants of the main station in Freiburg, Germany (all obtained from OSM data).
In general, the following characteristics of station labels can be found in the western world:
(1) Typos are rare. Station labels are usually very short and often used in information systems on the train or the stations, where typos would be noticed immediately.
(2) There are typical (but often regionally specific) abbreviations, like ``Str.'' or ``St.'' for ``street'' in German or English, respectively, or the ubiquitous ``Hbf'', short for ``Hauptbahnhof'' (main station), in Germany.
(3) Location specifiers like town, district or area names (e.g. the Breisgau area in Figure~\ref{FIG:labelfreiburg}) may be (partially) omitted. For example, in a schedule published by a city's local transit authority, it is not necessary to prefix every station label with the name of the city: a label \Pstation{Central Station} is usually a unique identifier inside municipal boundaries.
(4) Exact station labels may be (partially) omitted. For example, in long-distance train schedules, where trains only stop at a single station in town, a dropped suffix ``Central Station'' will not lead to confusion - it is enough to just give the name of the city.
(5) Token ordering may vary greatly.
(6) Station labels may exist in an official, full-length form (\Pstation{St Pancras International Station}) and in a colloquially used shorter form (``St Pancras'').
(7) Token separators vary greatly, and may - interchangeably - consist of whitespace, commas, semicolons, hyphens, brackets, or are indicated by camel casing (\Pstation{StPancras} instead of \Pstation{St~Pancras}, \Pstation{Freiburg(Breisgau)} instead of \Pstation{Freiburg~(Breisgau)}).
(8) labels may be over-specified and describe locations inside the station. For example, schedule data for a single railway line may explicitly mention the track number it usually arrives at.

\newcommand{\namebox}[2]{%
    \begin{tikzpicture}%
        \node [text height=1.7mm,minimum height=3.74mm,fill=#2, outer sep=0, inner sep=2, rounded corners=3pt, fill opacity=0.2, text opacity=1] {\footnotesize #1};
    \end{tikzpicture}%
}

\begin{figure}[t]
\begin{minipage}[c]{.15\textwidth}
    \raggedright
    \namebox{Hauptbahnhof}{green}\\[2pt]
    \namebox{Freiburg}{red}\\[2pt]
    \namebox{Freiburg}{red} \namebox{Hauptbahnhof}{green}\\[2pt]
    \namebox{Freiburg}{red} \namebox{Hbf}{green}\\[2pt]
    \namebox{Freiburg}{red} \namebox{im Breisgau}{blue}\\[2pt]
    \namebox{Hauptbahnhof}{green} \namebox{Freiburg}{red}\\[2pt]
\end{minipage}
\hfill
\begin{minipage}[c]{.25\textwidth}
    \namebox{Freiburg}{red} \namebox{im Breisgau}{blue} \namebox{Hauptbahnhof}{green}\\[2pt]
    \namebox{Freiburg}{red} \namebox{(Breisgau)}{blue} \namebox{Hauptbahnhof}{green}\\[2pt]
    \namebox{Hauptbahnhof}{green}\namebox{,}{cyan} \namebox{Freiburg}{red} \namebox{im Breisgau}{blue}\\[2pt]
    \namebox{Freiburg}{red} \namebox{(Breisgau)}{blue}\namebox{,}{cyan} \namebox{Hauptbahnhof}{green}\\[2pt]
    \namebox{Freiburg}{red}\hspace{-3.5pt}\namebox{(Brsg)}{blue} \namebox{Hauptbahnhof}{green}\\[2pt]
    \namebox{Freiburg}{red}\hspace{-3.5pt}\namebox{(Brsg)}{blue} \namebox{Hbf}{green}
\end{minipage}
    \vspace{-.45em}
    \caption{Incomplete list of different label variants in OSM for the main station (German: ``Hauptbahnhof'') in Freiburg. The location specifier ``im Breisgau'' is sometimes used to distinguish the town from Freiburg im \"Uechtland in Switzerland. Similar tokens are highlighted by the same color.}
    \label{FIG:labelfreiburg}
    \vspace{-1em}
\end{figure}

\section{Classification Techniques}
\label{SEC:classification}

In this section, we will describe several similarity classification techniques based on geographic coordinates, the station labels or combinations thereof.
We will first discuss two naive baseline approaches based on station label or station position equivalency in Section~\ref{SEC:baseline}.
We will then extend the latter to a similarity measure using a distance threshold in Section~\ref{SEC:possimi}.
After that, Section~\ref{SEC:labelsim} describes several methods to measure station label similarity, most of which are based on established string similarity measures.
In Section~\ref{SEC:combinations}, we combine these similarity measures.
In Section~\ref{SEC:ml}, we develop a machine learning based approach to our problem.

\subsection{Naive Techniques}
\label{SEC:baseline}

A naive solution to our classification problem would consider two station identifiers as equivalent if their positions and/or labels are equivalent.
It is already clear from Section~\ref{SEC:characteristics} that such an approach will perform very badly and lead to many false negatives.
Nevertheless, we describe both techniques, also as two simple examples for the formalism we use to describe all our techniques.

\subsubsection{Position Equivalency}

To decide whether two station identifiers $s_a$ and $s_b$ are similar, we simply use a function $c_{\scriptscriptstyle \text{PEQ}}(s_a, s_b)$ that checks whether their positions are equivalent (within some $\epsilon$ to account for floating point inaccuracies):
\begin{equation}
    c_{\scriptscriptstyle \text{PEQ}}(s_a, s_b) = \begin{cases}
        1, & \text{if } d\left(\phi_a, \lambda_a, \phi_b, \lambda_b\right) < \epsilon\\
        0, & \text{otherwise,}
    \end{cases}
\end{equation}
where $d$ is the geographic distance between $(\phi_a, \lambda_a)$ and $(\phi_b, \lambda_b)$.

\subsubsection{Label Equivalency}

Likewise, we define a function $c_{\scriptscriptstyle \text{LEQ}}$ that decides that $s_a$ and $s_b$ are similar if their labels are equivalent:
\begin{equation}
    c_{\scriptscriptstyle \text{LEQ}}(s_a, s_b) = \begin{cases}
        1, & \text{if } n_a = n_b\\
        0, & \text{otherwise.}
    \end{cases}
\end{equation}
This approach is not robust against small name deviations (for example, \Pstation{London St Pancras} vs. \Pstation{London St. Pancras}) stations in different cities sharing the same name.

\subsection{Station Position Similarity}
\label{SEC:possimi}

We may improve $c_{\scriptscriptstyle \text{PEQ}}$ from above by replacing $\epsilon$ with a distance threshold $\hat d$ under which two station identifiers are considered similar.
However, such a binary function would be hard to combine with other approaches (for example, using soft voting).
Instead, we would like to have a continuous score of whether two station identifiers are similar.
Based on the observation that distances between similar stations seem to follow an exponential distribution (Fig.~\ref{FIG:disthisto}), we model this as follows:
\begin{equation}
    \psimi(s_a, s_b) = \exp\left(\frac{-\ln(2) \cdot d\left(\phi_a, \lambda_a, \phi_b, \lambda_b\right)}{\hat d}\right),
\end{equation}
$d$ is again the geographic distance.
The rate parameter is set to $\ln(2)$ to ensure a median of 1 ($\psimi < 0.5$ when $d$ is bigger than $\hat d$).

\subsection{Station Label Similarity}
\label{SEC:labelsim}

To make the label comparison more robust, this section discusses several techniques to  measure station label similarity.

\subsubsection{Name Normalization}
\label{SEC:normalization}

Text normalization describes the process of canonizing input texts before they are further processed.
In the context of station labels, some of the differences in spelling and representation described in Section~\ref{SEC:characteristics} may be removed by manually created normalization rules.
As station labels are written in uppercase in some datasets, it may also be useful to transform all characters of a station label into upper- or lowercase letters.
Figure~\ref{FIG:norm_rules} gives examples of station label normalization rules for the German language (all operating on lowercase labels).

Another frequently used technique in text normalization is the concept of stop words.
Here, a list of words that are irrelevant for similarity is compiled either by hand or automatically.
For example, if the dataset only consists of stations from the public transit network of Berlin, it may be reasonable to assume that ``Berlin'' has no relevance for the similarity of station labels (as many stations will be prefixed with it).

\begin{figure}
    \footnotesize
    \fbox {%
    \begin{minipage}[t]{.15\textwidth}%
        \begin{align*}
            \texttt{,} &\longrightarrow \texttt{\textvisiblespace}\\
            \texttt{-} &\longrightarrow \texttt{\textvisiblespace}\\
            \texttt{"} &\longrightarrow \texttt{\textvisiblespace}\\
            \texttt{\&} &\longrightarrow \texttt{und}\\
            \texttt{+} &\longrightarrow \texttt{und}\\
            \texttt{\"a} &\longrightarrow \texttt{ae}\\
            \texttt{\ss} &\longrightarrow \texttt{ss}
        \end{align*}
    \end{minipage}%
    \begin{minipage}[t]{.312\textwidth}%
        \begin{align*}
            \texttt{str.} &\longrightarrow \texttt{strasse}\\
            \texttt{([a-z])strasse(\$|\texttt{\textvisiblespace})} &\longrightarrow \texttt{\textbackslash 1\textvisiblespace strasse\textbackslash 2}\\
            \texttt{st} &\longrightarrow \texttt{strasse}\\
            \texttt{(\textasciicircum|\texttt{\textvisiblespace})hbf\textbackslash.(\$|\texttt{\textvisiblespace})} &\longrightarrow \texttt{\textbackslash 1hauptbahnhof\textbackslash 2}\\
            \texttt{(\textasciicircum|\texttt{\textvisiblespace})hbf(\$|\texttt{\textvisiblespace})} &\longrightarrow \texttt{\textbackslash 1hauptbahnhof\textbackslash 2}\\
            \texttt{\textbackslash s+} &\longrightarrow \texttt{\textvisiblespace}\\
            \texttt{\textasciicircum\textbackslash s} &\longrightarrow\\
            \texttt{\textbackslash s\$} &\longrightarrow
        \end{align*}
    \end{minipage}}
    \caption{Excerpt of the manually compiled station label normalization rules used in our evaluation to measure the extent to which our classification approaches are robust against variants in spelling. Given as regular expressions (\texttt{\textbackslash<n>} matches the $n$-th matched group on the left hand side). All labels are transformed to lowercase first.}
    \label{FIG:norm_rules}
\end{figure}

\subsubsection{String Similarity Measures}

To our knowledge, the applicability of classic string similarity measures to station labels has not been investigated so far.
In our experiments, we will evaluate several well-known measures, briefly summarized in this section.

The classic edit distance $ed(s_a, s_b)$ counts the number of edits (add, delete or substitution) necessary to transform $s_a$ into $s_b$ \cite{levenshtein66}.
It can be transformed into a similarity measure by taking the ratio between the distance and length of the larger input string.

To make the similarity measure more robust against missing parts, many measures have been proposed \cite{navarro01, cohen03}.
One natural approach is to use the prefix edit distance (PED); it is defined as $ped(a, b) = \min_{b'} ed(a, b')$,
where $b'$ is a prefix of a \cite{DBLP:journals/tois/BastC13}.
As the PED is not symmetric, we compute the PED similarity in both directions and simply take the best result.

A robust similarity measure targeted especially at shorter strings is the Jaro similarity \cite{jaro89}.
The closely related Jaro-Winkler similarity favors strings which match from the beginning \cite{winkler90}.

The Jaccard index is a similarity measure based on the string tokens $A$ and $B$ of strings $s_a$ and $s_b$, respectively.
As the Jaccard index is not very robust against minor differences in spelling or missing tokens, we additionally evaluate a similarity score that aims to combine the advantage of the Jaccard index (ordering is irrelevant) and the edit distance similarity.
We call this measure the best token subsequence similarity (BTS) and define it as
\begin{align}
    \text{sim}_{\text{BTS}}(s_a, s_b) = \max\left(\max_{a \in P(A)} \mathit{sim}^{*}_{\text{ED}}(a, n_b), \max_{b \in P(B)} \mathit{sim}^{*}_{\text{ED}}(b, n_a) \right),
\end{align}
where $\mathit{sim}^{*}_{\text{ED}}$ is the edit distance similarity directly on strings.
$P(S)$ is the set of all possible permutations of all subsets of $S$ with size $1 \leq n \leq |S|$,
concatenated with a space.
For example,
\begin{align*}
P\left(\left\{\text{\Pstation{Freiburg}}, \text{\Pstation{Hauptbahnhof}}\right\}\right) = &\{\text{\Pstation{Freiburg}}, \text{\Pstation{Hauptbahnhof}},\\
&\text{\Pstation{Freiburg Hauptbahnhof}},\\
&\text{\Pstation{Hauptbahnhof Freiburg}}\}.
\end{align*}
Because $|P(S)|$ grows super-exponentially, the calculation cost for labels with many tokens is an obvious drawback.
In our experiments, we fall back to the Jaccard index if $|P(A)| > 6$ or $|P(B)| > 6$.

We also evaluate TFIDF scores, a standard method in Information Retrieval \cite{leskovec14}.
TFIDF scores are based on the term frequency (the number of times a token appears in a \emph{document}), and the document frequency (the number of documents a token occurs in).
They are calculated per token (in our case, documents are the labels itself).
As a similarity measure between two sets of string tokens $A$ and $B$, we then simply take the cosine similarity of their relevance vectors.

\subsection{Combined Techniques}
\label{SEC:combinations}

Classifiers based on label similarity tend to produce false positives, as stations in different cities often share a common name.
Conversely, classifiers based on geographic positions may fail if stations have a distance greater than the threshold, or produce false positives if two non-similar stations are positioned very close to each other.
A simple idea is to combine them.

However, the label similarity measures described above all give values between 0 and 1, and require some threshold $t$ for classification.
To make it easier to combine these measures, we would again like to have a continuous value that is 1 if the similarity measure is 1, 0.5 if the similarity measure is exactly $t$ and 0 if the similarity measure is 0.
For a given similarity measure and two station identifiers $s_a$ and $s_b$, we define $\mathit{sim}'$ like this:
\begin{align}
    \mathit{sim}'(s_a, s_b) = \begin{cases}
        \frac{1}{2} + \frac{\mathit{sim}(s_a, s_b) - t}{2 (1 - t)}& \text{if }\mathit{sim}(s_a, s_b) > t\\
        \frac{\mathit{sim}(s_a, s_b)}{2 t} & \text{otherwise.}
    \end{cases}
\end{align}
For example, if $\edsimi(s_a, s_b) = 0.9$ and $t = 0.8$, $\mathit{sim}'_{\text{ED}}(s_a, s_b) = 0.75$.
Using this, we define a function $c_{\mathit{sim}}$ such that $c_{\mathit{sim}}(s_a, s_b) = 1$ if $\mathit{sim}'(s_a, s_b) > 0.5$, or else $c_{\mathit{sim}}(s_a, s_b) = 0$.

We can then combine different thresholded similarity scores with a soft or hard voting approach.
In soft voting, the similarity scores given by the respective classifiers are averaged.
In hard voting, the final similarity score is calculated by a majority vote.

\subsection{Machine Learning}
\label{SEC:ml}

TFIDF scores already ``learn'' label tokens of low significance.
For example, in a dataset of London, the token \Pstation{London} would have low significance because of its high document frequency.
However, in a different area of Great Britain, \Pstation{London Street} may be a unique station label.
None of our classifiers so far considered this.

Additionally, there may be abbreviations which are either regionally specific or difficult to capture in classic similarity measures.
In Section~\ref{SEC:normalization}, we described label normalization by manually created rules.
The goal of this section is to build a classifier which can learn abbreviations and the regional specificity of tokens automatically.
We base our classifier on an off-the-shelf random forest classifier \cite{breiman01}, chosen for its ease of use and robustness.
\subsubsection{Feature Engineering}
\label{SEC:features}
\begin{figure}
    \centering
    \includegraphics[trim=0 10 0 0, clip, width=0.475\textwidth]{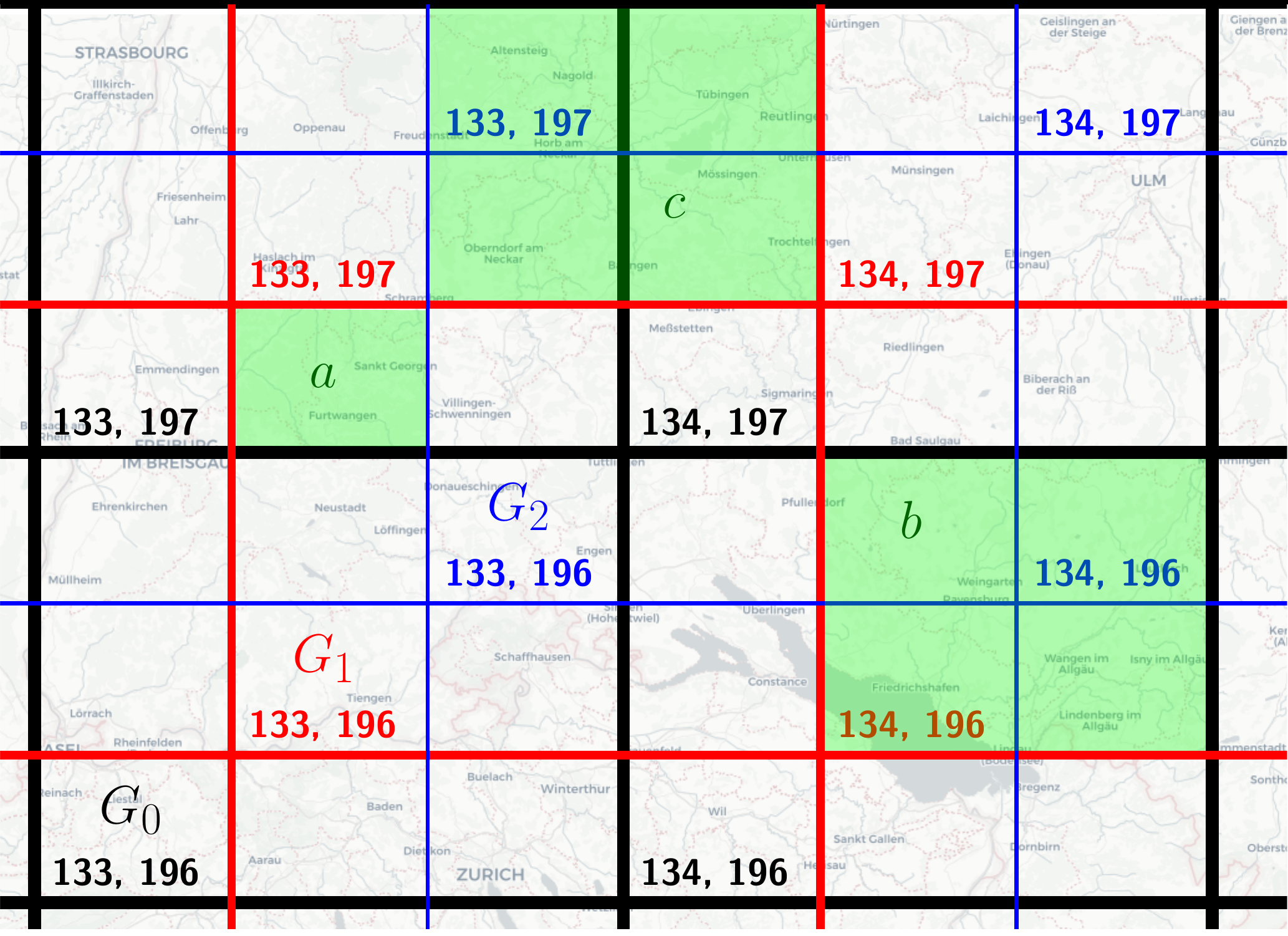}
    \vspace{-0.5em}
    \caption{Three interwoven grids $G_0$, $G_1$ and $G_2$ used to assign station identifiers to specific areas. Rectangle $a$ is uniquely identified by grid coordinates $g_0 = (133, 197), g_1 = (133, 196), g_2 = (132, 196))$, rectangle $b$ is uniquely identified by grid coordinates $g_0=(134, 196), g_1 = (134, 196)$. There is no selection of grid coordinates to identify $c$.}
    \vspace{-0.5em}
    \label{FIG:grid}
\end{figure}
\renewcommand\cellalign{cl}
\begin{table*}
  \caption[]{Example feature vectors for three station pairs in a testing dataset for the Freiburg area: (1) \Pstation{Freiburg im Breisgau Hauptbahnhof} @ $(47.9966, 7.8404)$ vs. \Pstation{Hauptbahnhof} @ $(47.9965, 7.8407)$. (2) \Pstation{Okenstra\ss e} @ $(48.0105, 7.8545)$ vs. \Pstation{Nordstra\ss e} @ $(48.0111, 7.8541)$.  (3) \Pstation{ZOB} @ $(47.9959, 7.8405)$ vs. \Pstation{Zentraler Omnibusbahnhof, Freiburg im Breisgau} @ $(47.9960, 7.8407)$. The distance in meters is given by $d_m$ and $d_{3g}$ is the number of trigrams that only occur in one of the two labels. Their relationship in terms of the top-15 trigrams is given by the absolute difference in occurrences between the two labels. $(x_0, y_0)$ and $(x_1, y_1)$ are the coordinates of the station pair centroid on two interwoven grids $G_0$ and $G_1$. \label{TBL:featvec}}
  \vspace{-3mm}
  \centering
  {\renewcommand{\baselinestretch}{1.0}\normalsize
  \setlength\tabcolsep{4.25pt}
  \begin{tabular*}{\textwidth}{l r r r r r r c c c c c c c c c c c c c c c c}
                 \# & $d_m$ & $d_{3g}$ & $x_0$ & $y_0$ & $x_1$ & $y_1$ & \ngram{rei} & \ngram{tra}  & \ngram{ra\ss}  & \ngram{a\ss e}  & \ngram{urg}  & \ngram{bur} & \ngram{ibu}  & \ngram{\textvisiblespace Fr}  & \ngram{Fre}  & \ngram{eib}  & \ngram{rg\textvisiblespace}   & \ngram{eis}   & \ngram{Bre}   & \ngram{sga}   & \ngram{isg}     & ``similar''\\\toprule    
    1 & 24 m & 20 & 133 & 196 & 133 & 195 & -2 & 0 & 0 & 0 & -1 & -1 & -1 & -1 & -1 & -1 & -1 & -1 & -1 & -1 & -1 & \textbf{yes} \\
    \midrule
    2 & 72 m & 10 & 133 & 196 & 133 & 195 & 0 & 0 & 0 & 0 & 0 & 0 & 0 & 0 & 0 & 0 & 0 & 0 & 0 & 0 & 0 & \textbf{no} \\
    \midrule
    3 & 12 m & 47 & 133 & 196 & 133 & 195 & 2 & 1 & 0 & 0 & 1 & 1 & 2 & 1 & 1 & 1 & 1 & 1 & 1 & 1 & 1 & \textbf{yes} \\
    \bottomrule
  \end{tabular*}}
\end{table*}
The classifier is trained on features of matching and non-matching station identifier pairs.
Table~\ref{TBL:featvec} gives an example of two feature vectors for data based on the OpenStreetMap data of the Freiburg region.
We use the following features:

(1) The meter distance between the two station identifiers.
    We want to give the model the possibility to learn to ignore deviations in station labels if the identifiers are close, and that high distances make it very unlikely that two stations are similar.

(2) The grid coordinate of the centroid of both station positions on an interwoven grid.
    We assume that the centroid is representative for the general area of the stations (Section~\ref{SEC:evalsetup} will make it clear that the distance between the two stations is always small enough for that to be the case).
    For $n$ interwoven grids $G_0, G_i, ..., G_n$ with grid cells of width $w$ and a height $h$, we offset the $x$ origin of each $G_i$ by $w/n$, and the $y$ origin by $h/n.$
    Figure~\ref{FIG:grid} gives an example of such an interwoven grid with $n = 3$.
    The motivation behind this is to soften the effect of hard grid boundaries, and to also give the model the ability to learn about rectangular areas of varying sizes (for example, cell $a$ in Figure~\ref{FIG:grid} can be uniquely identified by a triplet of coordinates on all three grids).

(3) The difference in the number of occurrences of the training dataset's top $k$ trigrams between the right-hand side station label and the left-hand side station label.
    We take the trigrams from the original station labels, padded with a single space on both sides.
    For example, the trigrams for \Pstation{London} are \ngram{\textvisiblespace Lo}, \ngram{Lon}, \ngram{ond}, \ngram{ndo}, \ngram{don} and \ngram{on\textvisiblespace}.
    The padding makes sure that single character tokens are always represented by a distinct trigram.
    Table~\ref{SEC:features} gives an example how the differences are then calculated.
    For example, if \ngram{rei} occurs 2 times in the left-hand side station label, and 0 times in the right station label, the difference is -2. If \ngram{tra} occurs 1 time on the left-hand side label, and 1 time on the right-hand side label, the difference is 0.
    These features act as a simple language model.

(4) The number of non-matching trigrams (every trigram, not just the top $k$) between the station labels.
    If $A_3$ is the set of trigrams in station label $n_a$, and $B_3$ the set of trigrams in station label $n_b$, this number is given by $|A_3\cup B_3| - |A_3\cap B_3|$.
    The primary motivation for this feature is to capture the difference between station labels which do not contain any of the top $k$ trigrams.
    We use an absolute number here and not, for example, the Jaccard similarity because we want to enable the model to learn that a high number of missing trigrams is acceptable if the difference for a trigram of low significance accounts for it.

\section{Evaluation Setup}
\label{SEC:evalsetup}

\begin{figure}
    \centering
    \includegraphics[trim=0 35 0 40, clip, width=0.48\textwidth]{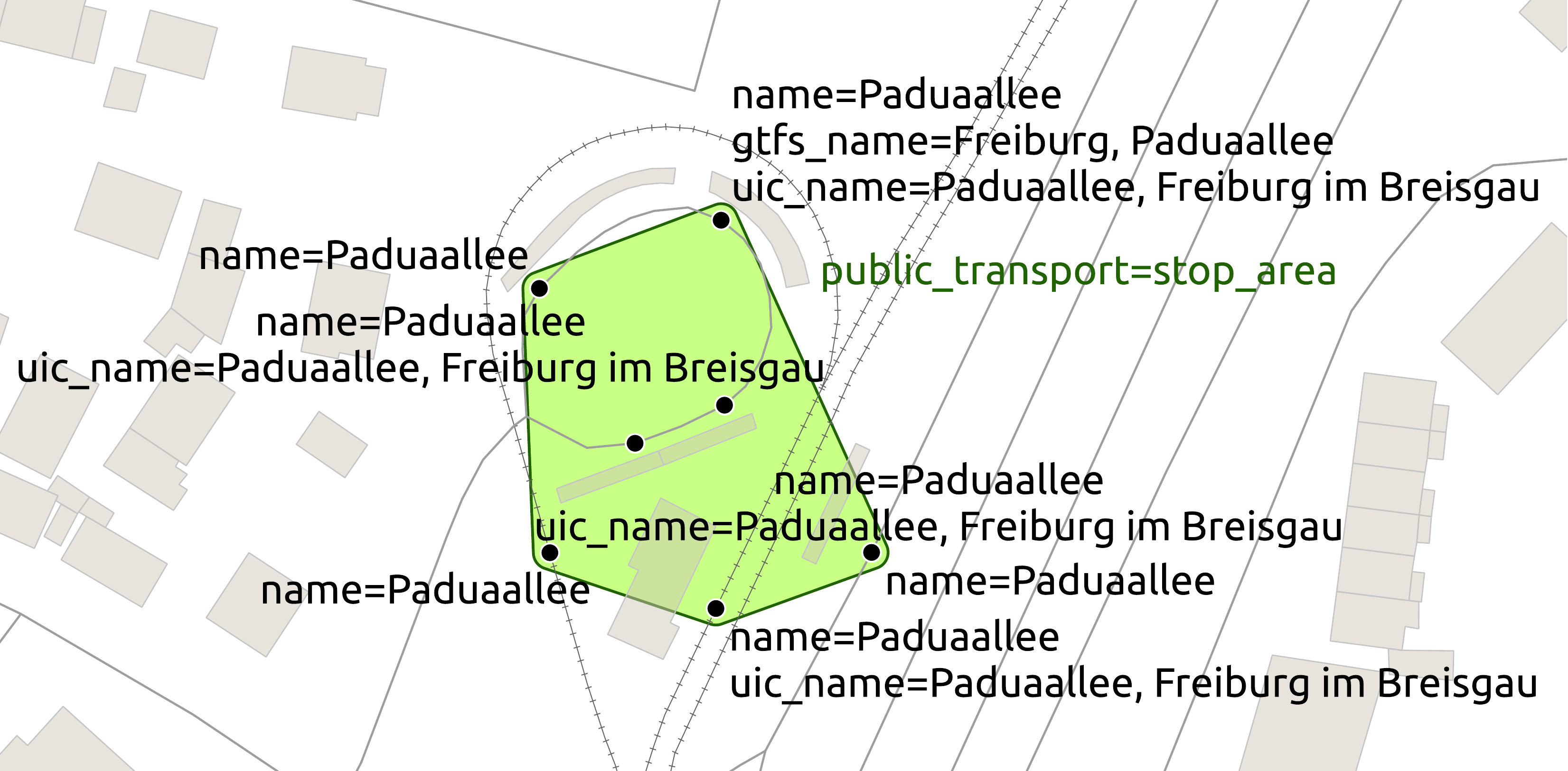}
    \caption{Typical bus/tram station in OpenStreetMap. Multiple stop nodes, each with possibly multiple labels (\texttt{name}, \texttt{uic\_name}, \texttt{ref\_name}, \texttt{gtfs\_name}, ...), are (manually) grouped by a \texttt{public\_transport=stop\_area} relation.}
    \label{FIG:osmpadua}
\end{figure}

As we know of no comprehensive station dataset that contains both typical spelling variants of station labels and also integrates different placement philosophies, we build our ground truth datasets from OpenStreetMap (OSM) data.
Figure~\ref{FIG:osmpadua} gives an example of a typical bus/tram station as it appears in OSM.
In the context of this work, we only consider station objects tagged as nodes.
Polygonal stations (buildings, platforms) are not used, although they can easily be included in our approach.
Stations in OSM often come with multiple label attributes.
For example, \texttt{name=*} gives a generic label and \texttt{reg\_name=*} sometimes contains a regionally used label.
Table~\ref{TBL:name_attrs} lists the station label attributes we use.
The station nodes may be grouped by a relation \texttt{public\_transport=stop\_area}, which again may come with one or multiple label attributes\footnote{\url{https://wiki.openstreetmap.org/wiki/Tag:public_transport}}.

From this data, we build our ground-truth data like this: each station label yields a station identifier with the position of the station node.
If the enclosing \texttt{public\_transport=stop\_area} relation contains labels not present in a station node's labels, we add them to the node.
We then count a pair $\{s_a, s_b\}$ of station identifiers, where both $s_a$ and $s_b$ are inside the same \texttt{stop\_area} relation, as ``similar''.
A pair $\{s_a, s_b\}$ of station identifiers, where $s_a$ is inside a \texttt{stop\_area} relation $A$, and $s_b$ is in \emph{another} \texttt{stop\_area} relation $B$, is marked as ``not similar''.
Station nodes not contained in a \texttt{stop\_area} relation (``orphan'' nodes) are never marked as ``not similar'' to anything (but station identifiers generated from their labels are pairwise marked as ``similar''), as station objects are sometimes forgotten to be included in \texttt{stop\_areas}.
Note that ignoring these orphan nodes does not select an ``easy'' subset of the data.
Stops without a \texttt{stop\_area} relation are usually very simple cases in rural areas (two stops on opposite sides of the road, sharing the same single label).

We additionally apply two heuristics to keep our ground truth clean: (1) if two station identifiers are not in the same \texttt{stop\_area}, but have exact matching names and are within 250 meters, we ignore this pair.
If two station identifiers are in different \texttt{stop\_area} relations, but the relations themselves are grouped by a super-relation \texttt{public\_transport=stop\_area\_group}, we also ignore this pair.

To avoid an unnecessarily large number of ``not similar'' pairs, we set a search radius threshold.
Above this threshold, we implicitly assume that a pair can always by trivially considered as ``not similar''.
In this work, we used a threshold of 1,000 meters.
For both our ground truth datasets, the original input data did not contain similar station identifier pairs with a distance over 1,000 meters (except for a few mapping mistakes), so no interesting cases were lost.

\renewcommand\cellalign{cl}
\begin{table}
  \caption[]{Name attributes for station nodes in OpenStreetMap (OSM) used in our ground truth dataset. \label{TBL:name_attrs}}
  \vspace{-3mm}
  \centering
  {\renewcommand{\baselinestretch}{1.0}\normalsize
  \setlength\tabcolsep{3pt}
  \begin{tabular*}{.48\textwidth}{l l}
                 attribute & description \\\toprule
    \texttt{name} & Generic label used by default. \\
    \midrule
    \texttt{ref\_name} & Sometimes gives a fully-qualified label. \\
    \midrule
    \texttt{uic\_name} & UIC label, often equivalent to \texttt{ref\_name}. \\
    \midrule
    \texttt{official\_name} & Often equivalent to \texttt{ref\_name}. \\
    \midrule
    \texttt{alt\_name} & An alternative label. \\
    \midrule
    \texttt{loc\_name} & Local station label (without location specifier). \\
    \midrule
    \texttt{reg\_name} & Regional label (without location specifier). \\
    \midrule
    \texttt{short\_name} & Short label. \\
    \midrule
    \texttt{gtfs\_name} & \makecell{Undocumented, sometimes states the label used\\ in local schedule data.} \\
    \bottomrule
  \end{tabular*}}
\end{table}

\subsection{Spicing}
\label{SEC:spicing}

Two station identifiers labeled ``London St Pancras'' and ``Berlin Hauptbahnhof'' are obviously not similar, even if they are positioned only a few meters away.
However, such mapping mistakes would quickly be fixed by the OSM community and thus not appear in our ground truth.

For our ground truth to better match real-world input data, we randomly add such station pairs.
We refer to this process as \emph{spicing}.
Namely, for each original station identifier $s_a$, we select with probability $p$ a random set of 5 station identifiers $s_b^1, ..., s_b^5$ outside of the search radius. Each $s_b^i$ is given a random coordinate within 100 meters of $s_a$ and $\{s_a, s_b^i\}$ added as a ``not similar'' pair.

To simulate coordinate imprecision which is often present in real-word datasets (as discussed in Section~\ref{SEC:poscharacteristics}), we select with probability $p$ a similar station pair and add gaussian noise (with a standard deviation of 100 meters) to the coordinates of one station.

The effect of this spicing on the general performance of our classifiers will be evaluated in Section~\ref{SEC:norm}.

\section{Experimental Results}
\label{SEC:results}

\def\Hghost{\makebox[1.3mm][l]{}}
\def\Hk{\makebox[1.3mm][l]{\hspace{0.2mm}\footnotesize k}}
\def\HM{\makebox[1.3mm][l]{\hspace{0.2mm}\footnotesize M}}
\renewcommand\cellalign{cl}
\begin{table}
  \caption[]{Dataset dimensions for Great Britain and Island (BI) and Germany, Austria and Switzerland (DACH).
  $N$ is the number of stations, $G$ the number of groups, $N'$ the number of stations without a group (orphan stations), $|s|$ the number of unique station identifiers, $g$ the average group size, $d^+$ the average meter distance between positive ground truth pairs, $K^-$ the number of ``not similar'' pairs and $K^+$ the number of ``similar'' pairs (all without spicing).\label{TBL:dataset_sizes}}
  \vspace{-3mm}
  \centering
  {\renewcommand{\baselinestretch}{1.0}\normalsize
  \setlength\tabcolsep{3.6pt}
  \begin{tabular*}{.48\textwidth}{l r r r r r r r r r}
                 & $N$ & $G$ & $N'$ & $|s|$ & $g$ & $d^+$ & $K^-$ & $K^+$ & \textbf{\emph{K}} \\\toprule
    BI & 270\Hk& 15\Hk & 234\Hk & 261\Hk & 3.7 & 56.7 & 1.7\HM & 0.4\HM & \textbf{2.1\HM}\\
    \midrule
    DACH & 679\Hk & 102\Hk & 350\Hk & 875\Hk & 5 &  46.1 & 11.1\HM & 2.6\HM & \textbf{13.6\HM}\\
    \bottomrule
  \end{tabular*}}
\end{table}

We evaluated two datasets: the OSM data for the British Isles (Great Britain and Ireland, BI) and the OSM data for Germany, Austria and Switzerland (DACH).
The latter yielded over 13 million station identifier pairs.
Their exact dimensions are given in Table~\ref{TBL:dataset_sizes}.

Our interest was twofold: first, we wanted to find out whether simple classification methods based on similarity measures have a natural cutoff below which two stations can be considered not similar, and above which they can be considered similar.
This was motivated by the fact that in real-world applications, a heuristic cutoff value for some similarity measure is usually employed to determine station similarity.
Second, we wanted to compare the best possible performance of each simple classification method against our machine-learning based method.

To this end, we determined the optimal threshold values for each similarity measure classifier $c_{\mathit{sim}}$ (when used in a standalone fashion) described in Section~\ref{SEC:classification}:
geographical distance (P), edit distance (ED), prefix edit distance (PED), Jaro similarity (J), Jaro-Winkler similarity (JW), Jaccard index (JAC), best-token subsequence similarity (BTS), and TFIDF similarity (TFIDF).
Afterwards, we evaluated combinations of those classifiers (P + ED, P + BTS, and P + TFIDF).
In Section~\ref{SEC:eval_rf}, we discuss the performance of our random forest classifier (RF) in more detail.
We measure the effect of the number of used top-$k$ trigrams and discuss the effects of a more fine-grained interwoven geographic grid.

All classifiers were evaluated in terms of the number of true positives TP, the number of true negatives TN, the number of false positives FP and the number of false negatives FN.
We evaluated precision, recall and F1 scores.
Precision scores were calculated as $\frac{\text{TP}}{\text{TP}+\text{FP}}$, recall scores as $\frac{\text{TP}}{\text{TP}+\text{FN}}$.
The F1 score is the harmonic mean between precision and recall.

For the evaluation, the ground truth dataset was spiced (see Section~\ref{SEC:spicing}) with probability $p=0.5$.
We then divided the ground truth data into a training set (a random selection of 20\% of the ground truth data) and a test set (the remaining 80\%).
There are several reasons why we opted for this unusual ratio:
    (1) Only our TFIDF and RF classifiers required an actual training step, and we wanted to evaluate all classifiers against the same test dataset. A bigger training dataset (70 - 80\% of the ground truth data) would have required us to limit our evaluation of the baseline approaches to only a small fraction of our datasets.
    (2) Because of the high quality of the OSM data, our ground truth data was extensive. There was no need to restrict the evaluation to a small sample size to gain more training data.
    (3) We were interested in the performance of the RF classifier when trained on only a small sample of the ground truth dataset.
    (4) Faster training.

For the RF classifier, we also experimented with bigger training datasets (80\% of the ground truth), but found the performance gain to be minimal (note that when trained on 20\% of the ground truth, the F1 score of the RF classifier is already above 99\%).
Classifiers that didn't require training were evaluated against the test dataset.
Classifiers that required training (TFIDF and RF) were trained on the training dataset, and evaluated against the test dataset.

All evaluation runs were repeated 5 times (each time with a randomly divided test/training dataset that was the same for all classifiers) and final scores averaged.
An overview of our results is given in Table~\ref{TBL:evalres}.
The evaluation setup can be found online\footnote{\url{https://github.com/ad-freiburg/statsimi-eval}}.

A major takeaway of our experiments is that for a typical station identifier dataset such as ours, there is a large percentage (between 90 and 95\%) of cases which seem to be very easy to classify correctly using a fixed cutoff value.
The remaining $5$--$10$\%, however, are hard to crack with conventional similarity measures or combinations thereof, even when we search for the optimal cutoff values beforehand.
Interestingly, the optimal cutoff values for the geographic and the edit distance were the same for both the DACH and BI dataset (when used as standalone classifiers), suggesting that these values might be language and area independent.
This was also the case when geographic and edit distance were combined.

For the DACH dataset, we additionally evaluated the effect of manual station label normalization for all classifiers in Section~\ref{SEC:norm}.

\newcommand\PA[1]{\makebox[5.05mm][l]{#1}}
\newcommand\Ppairth[2]{\hspace{-.5em} #1\hspace{0.2em}+\hspace{0.2em}\PA{#2}\hspace{-.5em} }
\renewcommand\cellalign{cl}
\begin{table}
  \caption[]{Evaluation results for best parameters (optimized for best F1 score), on unnormalized, spiced input. The value(s) of the similarity measure thresholds is given by $t$.\label{TBL:evalres}}
  \vspace{-3mm}
  \centering
  {\renewcommand{\arraystretch}{0.98}\normalsize
  \setlength\tabcolsep{7.99pt}
  \begin{tabular}{l @{}l c r r r}
                 & method & t & prec. & rec. & F1\\\toprule
    \parbox[t]{7mm}{\multirow{12}{*}{\rotatebox[origin=c]{90}{BI}}}
         & P & \makebox[8.01mm][l]{100 m} & \PA{0.66} & \PA{0.93} & \PA{0.77}\\
         & ED & \PA{0.85} & \PA{0.99} & \PA{0.86} & \PA{0.92}\\
         & PED & \PA{0.85} & \PA{0.93} & \PA{0.89} & \PA{0.91}\\
         & J & \PA{0.9} & \PA{0.98} & \PA{0.86} & \textbf{\PA{0.92}}\\
         & JW & \PA{0.95} & \PA{0.99} & \PA{0.84} & \PA{0.91}\\
         & JAC & \PA{0.75} & \PA{0.99} & \PA{0.84} & \PA{0.91}\\
         & BTS & \PA{0.85} & \PA{0.91} & \PA{0.9} & \PA{0.91}\\
         & TFIDF & \PA{0.99} & \PA{0.99} & \PA{0.84} & \PA{0.91}\\
         \cmidrule{2-6}
         & P+ED & \Ppairth{40 m}{0.6} & \PA{0.96} & \PA{0.9} & \PA{0.93}\\
         & P+BTS & \Ppairth{10 m}{0.5} & \PA{0.93} & \PA{0.9} & \PA{0.91}\\
         & P+TFIDF & \Ppairth{150 m}{0.99} & \PA{0.96} & \PA{0.92} & \textbf{\PA{0.94}}\\
         \cmidrule{2-6}
         & RF & --- & $>$ \PA{0.99} & \PA{0.99} & $>$ \PA{\textbf{0.99}}\\
    \toprule
    \parbox[t]{7mm}{\multirow{12}{*}{\rotatebox[origin=c]{90}{DACH}}}
         & P & \makebox[8.01mm][l]{125 m} & \PA{0.4} & \PA{0.96} & \PA{0.56}\\
         & ED & \PA{0.85} & \PA{0.99} & \PA{0.67} & \PA{0.8}\\
         & PED & \PA{0.9} & \PA{0.93} & \PA{0.73} & \PA{0.82}\\
         & J & \PA{0.85} & \PA{0.93} & \PA{0.71} & \PA{0.8}\\
         & JW & \PA{0.9} & \PA{0.9} & \PA{0.72} & \PA{0.8}\\
         & JAC & \PA{0.45} & \PA{0.85} & \PA{0.88} & \PA{0.86}\\
         & BTS & \PA{0.85} & \PA{0.92} & \PA{0.93} & \PA{\textbf{0.92}}\\
         & TFIDF & \PA{0.7} & \PA{0.9} & \PA{0.85} & \PA{0.87}\\
         \cmidrule{2-6}
         & P+ED & \Ppairth{40 m}{0.55} & \PA{0.9} & \PA{0.83} & \PA{0.86}\\
         & P+BTS & \Ppairth{10 m}{0.6} & \PA{0.96} & \PA{0.89} & \PA{0.92}\\
         & P+TFIDF & \Ppairth{60 m}{0.5} & \PA{0.94} & \PA{0.93} & \PA{\textbf{0.94}}\\
         \cmidrule{2-6}
         & RF & --- & $>$ \PA{0.99} & $>$ \PA{0.99} & $>$ \PA{\textbf{0.99}}\\
    \bottomrule
  \end{tabular}}
  \vspace{-5pt}
\end{table}

\subsection{Similarity Measure Classifier Results}

\begin{figure}
    \begin{center}
        \resizebox {0.484\textwidth} {!} {
            \input{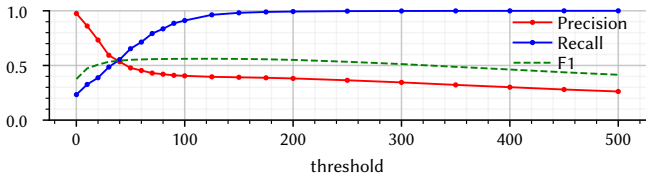}
        }
    \end{center}
    \vspace{-5pt}
    \caption{Effect of the threshold value (in meters) on precision, recall and F1 score for our geographic distance classifier (P) on the DACH dataset.}
    \label{FIG:t-p}
\end{figure}
\begin{figure}
    \vspace{-5pt}
    \begin{center}
        \resizebox {0.484\textwidth} {!} {
            \input{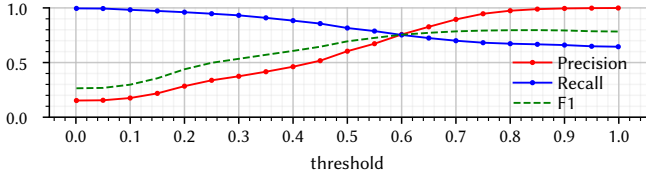}
        }
    \end{center}
    \vspace{-5pt}
    \caption{Effect of the threshold value on precision, recall and F1 score for our edit distance classifier (ED) on the DACH dataset.}
    \label{FIG:t-ed}
\end{figure}

For our classifiers based on geographic distance (P), edit distance similarity (ED), prefix edit distance similarity (PED), Jaro similarity (J), Jaro-Winkler similarity (JW), the Jaccard index (JAC), the best token subsequence edit distance (BTS), and TFIDF scores (TFIDF) we evaluated the threshold that maximized the classifier's F1 score.

For example, Figures~\ref{FIG:t-p} and \ref{FIG:t-ed} shows the effect of the threshold value on the geographic distance similarity (P) and edit distance similarity (ED) classifier for the DACH dataset.

For our DACH datasets, the station label similarity classifiers based on the best token subsequence similarity (BTS) and TFIDF scores performed best when used in a standalone fashion.
BTS was the clear winner among the standalone similarity measure based classifiers.
This is surprising, as TFIDF scores include an elaborate preprocessing step which tries to estimate the significance of certain tokens, and the BTS-based similarity scores operate completely locally on two station pairs.
However, on the BI dataset, there was only little variance (around 1\%) between the standalone string similarity measures.
A manual investigation showed that in Great Britain and Ireland, stations are much more consistently labeled in OSM than in the German speaking world.
This is demonstrated by the high F1 score of the ED classifier on the BI dataset (92\%).
One explanation for this is that in Germany, Austria and Switzerland, station objects in OSM often contain all different official labeling variants, while in Great Birtain and Ireland, there is often only a single, distinct label recorded.

\begin{figure}
    \begin{center}
        \begin{tabular*}{.48\textwidth}{l l}
            \makecell{\textbf{FN}} & \makecell{\Pstation{Parkweg} @ $(52.0149, 7.2051)$\\\Pstation{Rosendahl, Osterwick, Parkweg} @ $(52.0149, 7.2051)$} \\[10pt]
            \makecell{\textbf{FP}} & \makecell{\Pstation{Bruck an der Mur} @ $(47.4136, 15.2793)$\\\Pstation{Bruck an der Mur, Waldweg} @ $(47.4185, 15.2736)$}
        \end{tabular*}
    \end{center}
    \vspace{-1em}
    \caption{Typical false negative and false positive for a Jaccard index based classifier on our DACH dataset.}
    \label{FIG:false_jaccard}
    \vspace{-5pt}
\end{figure}
\begin{figure}
    \begin{center}
        \begin{tabular*}{.48\textwidth}{l l}
            \makecell{\textbf{FN}} & \makecell{\Pstation{Bromley-By-Bow Platform 2} @ $(51.5248, -0.0115)$\\\Pstation{Bromley By Bow Station} @ $(51.5234, -0.0121)$}  \\[10pt]
            \makecell{\textbf{FP}} & \makecell{\Pstation{Clapton Girls' Academy} @ $(1.5539, -0.0537)$\\\Pstation{Clapton} @ $(51.5617, -0.0568)$} 
        \end{tabular*}
    \end{center}
    \vspace{-1em}
    \caption{Typical false negative and false positive for a prefix edit distance based classifier on our BI dataset.}
    \label{FIG:false_ed}
    \vspace{-5pt}
\end{figure}
\begin{figure}
    \begin{center}
        \begin{tabular*}{.48\textwidth}{l l}
            \makecell{\textbf{FN}} & \makecell{\Pstation{Auerbach (Karlsbad), Rosenweg} @ $(48.9161, 8.5341)$\\\Pstation{Rosenweg} @ $(48.9160, 8.5343)$}  \\[10pt]
            \makecell{\textbf{FP}} & \makecell{\Pstation{Cottbus, Kiekebusch Alte Schule} @ $(51.7215, 14.3646)$\\\Pstation{Kiekebusch Friedhof, Cottbus} @ $(51.7179, 14.3672)$}
        \end{tabular*}
    \end{center}
    \vspace{-1em}
    \caption{Typical false negative and false positive for a TFIDF based classifier. On large datasets, TFIDF scores give tokens too little significance which are common nationally (like ``Schule'' (school) and ``Friedhof'' (cemetery)), but highly specific locally. Regionally common, but nationally rare tokens like the village name ``Auerbach'' near Karlsbad are given too much significance.}
    \label{FIG:false_tfidf}
    \vspace{-5pt}
\end{figure}
\begin{figure}
    \begin{center}
        \begin{tabular*}{.48\textwidth}{l l}
            \makecell{\textbf{FN}} & \makecell{\Pstation{Little Ilford School} @ $(51.5483, 0.0577)$\\\Pstation{Church Road} @ $(51.5479, 0.0569)$}  \\[10pt]
            \makecell{\textbf{FP}} & \makecell{\Pstation{Galsworthy Road/Moonshine Lane} @ $(53.4178, -1.4808)$\\\Pstation{Moonshine Lane - Galsworthy Road} @ $(53.4178, -1.4803)$}
        \end{tabular*}
    \end{center}
    \vspace{-1em}
    \caption{Typical false negative and false positive for our RF based classifier. FN: \Pstation{Little Ilford School} and \Pstation{Church Road} (in London) have not been grouped correctly in OSM; our model found a mapping mistake. FP: Different stations named after intersections of the same streets are often incorrectly marked as similar, because our RF classifier does not consider the ordering of trigrams.}
    \label{FIG:false_rf}
    \vspace{-5pt}
\end{figure}

When combined with a geographic distance based classifier (P), the F1 scores of ED, BTS and TFIDF generally improved.
For both our BI and DACH dataset, the best obtainable F1 score for such a classifier was 0.94 (P+TFIDF).

The evaluation results for our naive baseline techniques (station label equivalency or station position equivalency) on our DACH dataset can be read from Figures~\ref{FIG:t-p} and \ref{FIG:t-ed}.
The recall for station label equivalency was 0.67, and 0.22 for position equivalency.
Precision for position equivalency was nearly 1.0, which was to be expected, as there are basically no cases where different  stations share the same coordinate.
As our ground-truth data only considered non-similar stations up to a distance threshold of 1,000 meters, the precision of full name equivalency was also nearly 1.0.
The F1 score was 0.79 for label equivalency, and 0.39 for position equivalency.

\subsection{Random Forest Classifier Results}
\label{SEC:eval_rf}

\begin{figure}
    \begin{center}
        \resizebox {0.484\textwidth} {!} {
            \input{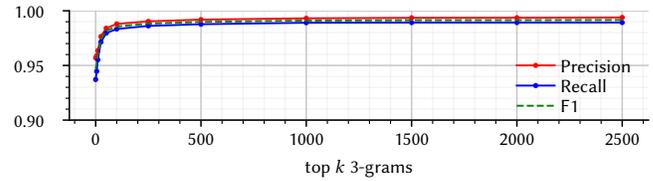}
        }
    \end{center}
    \vspace{-5pt}
    \caption{Effect of the number of top-$k$ trigrams on precision, recall and F1 score for our random forest (RF) classifier on the DACH dataset. The number of interwoven geographic grids stayed fixed at 2.}
    \label{FIG:t-rftopk}
\end{figure}

For our machine learning based approach, we used an off-the-shelf (from the Python scikit-learn library\footnote{\url{https://scikit-learn.org/}}) random forest (RF) classifier with default parameters (the number of trees was left at $100$).
For all our testing datasets, we used the top-$2500$ trigrams and 2 interwoven grids $G_0$ and $G_1$.
We did not use a separate validation set to optimize these hyperparameters, but used a different randomly selected training and testing dataset than in the evaluation.
The base grid cell dimensions are chosen in such a way that the earth is completely covered by a $256\times256$ grid (this means the cell width and height are around 156 km at the equator; conveniently, a single coordinate also fits into an 8 bit integer).
We evaluated other numbers for the top-$k$ trigrams and other numbers of grids.
For the number $k$ of top-$k$ trigrams, we found that the results quickly converge to the optimal F1 score.
For example, for our DACH dataset, the improvements for $k > 1000$ were marginal (Fig.~\ref{FIG:t-rftopk}).
Regarding the number of interwoven grids, we were surprised to find that the quality decreases after 2.
This may be explained with regional overfitting: a higher number of grids enables the encoding of smaller geographic areas.
Our classifier may then learn that certain location specifiers have little significance near an individual station, but may not generalize that this is also true for the greater surrounding area.

The RF classifier clearly outperformed every other classifier.
For both our datasets, precision and recall were at over 0.99.

After intensive manual investigation, we found four prevalent causes for the remaining false negatives and positives: (1) ambiguous cases where it is disputed whether stations belong to each other, (2) extreme outliers, e.g. similar identifiers that are more than 500 meters away and/or have highly abbreviated station labels, (3) different stations named after intersections of the same streets are incorrectly marked as similar (see FP example in Figure~\ref{FIG:false_rf}), (4) mapping mistakes in OSM.

\subsection{Impact of Spicing and Normalization}
\label{SEC:norm}
\renewcommand\cellalign{cl}
\begin{table}
    \caption[]{Effect of spicing and label normalization (with manually created rules) on our DACH dataset. Threshold values are again optimized for best F1 score. The percentages give the improvement compared to the best results without normalization and with spicing from Table~\ref{TBL:evalres}.\label{TBL:normalization}}
  \vspace{-1em}
  \centering
  {\renewcommand{\arraystretch}{1}\normalsize
  \setlength\tabcolsep{2.3pt}
  \begin{tabular}{@{}l l c c r r c c r r r}
         &&\multicolumn{4}{c}{without spicing} && \multicolumn{4}{c}{with normalization} \\
         \cline{3-6} \cline{8-11} \\[-2ex]
         method && t && F1 & impr. && t && F1 & impr. \\\cmidrule[\heavyrulewidth]{1-1}\cmidrule[\heavyrulewidth]{3-6}\cmidrule[\heavyrulewidth]{8-11} 
         P && \PA{125 m} && \PA{0.95}& +69.5\% && \PA{125 m} && \PA{0.56}& +0\% \\
         ED && \PA{0.85} && \PA{0.8}& +0\% && \PA{0.85} && \PA{0.81}& +1.3\% \\
         PED && \PA{0.85} && \PA{0.82} & +0.2\% && \PA{0.9} && \PA{0.83}& +1.1\% \\
         J && \PA{0.85} && \PA{0.81} & +0.1\% && \PA{0.65} && \PA{0.88}& +1.7\% \\
         JW && \PA{0.9} && \PA{0.8} & +0.2\% && \PA{0.95} && \PA{0.81}& +1.4\% \\
         JAC && \PA{0.45} && \PA{0.87} & +0.4\% && \PA{0.65} && \PA{0.88}& +1.7\% \\
         BTS && \PA{0.85} && \PA{0.93} & +0.2\% && \PA{0.95} && \PA{0.93}& +0.8\% \\
         TFIDF && \PA{0.65} && \PA{0.87} & -0.2\% && \PA{0.7} && \PA{0.87}& -0.2\% \\
         \cmidrule{1-1}\cmidrule{3-6}\cmidrule{8-11}
         P+ED && \Ppairth{50 m}{0.1} && \PA{0.97} & +12.1\% && \Ppairth{30 m}{0.55} && \PA{0.87}& +0.7\% \\
         P+BTS && \Ppairth{30 m}{0.1} && \PA{0.96} & +4\% && \Ppairth{10 m}{0.99} && \PA{0.95}& +2.7\% \\
         P+TFIDF && \Ppairth{50 m}{0.05} && \PA{0.96} & +2.5\% &&\Ppairth{60 m}{0.55} && \PA{0.94}& +0.3\% \\
         \cmidrule{1-1}\cmidrule{3-6}\cmidrule{8-11}
         RF && --- && $>$\PA{0.99} & +0.5\% && --- && $>$\PA{0.99} & +0.1\% \\
    \cmidrule[\heavyrulewidth]{1-1}\cmidrule[\heavyrulewidth]{3-6}\cmidrule[\heavyrulewidth]{8-11}
  \end{tabular}}
\end{table}
To measure the impact of normalization and the robustness of our techniques against a lack thereof, we re-ran the evaluation for our DACH dataset with prior normalization, using manually compiled rules as described in Section~\ref{SEC:normalization}.
The results are given in Table~\ref{TBL:normalization}, right column group.

The maximum F1 score improvement of 2.7\% for the similarity measure based classifiers was below our expectation.
For our RF classifier, the impact of manual normalization was minimal (around 0.1\%).
This indicates that the classifier learned these normalization rules during the training phase.
The TFIDF based classifiers also showed little to no improvement.
Tokens typically used in our normalization rules may already have a very high document frequency, limiting the impact of their normalization.

We note that our JAC, BTS and TFIDF classifiers already perform implicit normalization.
As these classifiers operate on word tokens, we have to choose some way of tokenization.
We are using a simple split by non-word characters, which effectively means that labels like \Pstation{St. Pancras} are normalized to \Pstation{St Pancras}.

Table~\ref{TBL:normalization} also gives the impact of the spiced station pairs we add to the ground truth (see Section~\ref{SEC:spicing} for details.)
The performance of classifiers based on geographic distance greatly improved if spicing was disabled.
This was to be expected, as the unspiced ground truth data from OSM is based on a curated dataset with few coordinate precision problems.

\section{Conclusions}
\label{SEC:conclusions}

We investigated how to automatically decide whether two station identifiers (each consisting of a label and a coordinate) belong to the same real-world station.
We discussed several approaches to this problem.
Our evaluation on extensive ground truth data obtained from OpenStreetMap (OSM) data showed that typical datasets have a large percentage (90--95\%) of ``easy'' cases where simple techniques based on edit or geographic distance may already perform well.
For practical use, however, they are not good enough, especially when a robustness against coordinate imprecisions is required (which is typically the case).
As expected, more elaborate similarity measures for station labels improved the overall classification performance.
However, on our biggest dataset (DACH), the best classifier based on a similarity measure still only achieved an F1 score of 94\%.
In contrast, our learning-based approach achieved F1 scores above $99\%$ across all datasets and even found errors in the original OSM data.

It might be of interest to further evaluate the robustness of our approach against spelling errors.
We would also like to better evaluate the extent to which our learning-based classifier is able to learn locally irrelevant location specifiers, for example by constructing a ground-truth dataset in which these specifiers can easily be separated from the station labels.
Such a dataset may be constructed from the administrative boundaries contained in the OSM data.

\bibliographystyle{ACM-Reference-Format}
\bibliography{simi}

\end{document}